\documentclass[aps,pre,twocolumn,showpacs]{revtex4}

\usepackage{dcolumn}
\usepackage{amsmath}
\usepackage{amssymb}
\usepackage{graphicx}

\begin{document}
\newcommand \be {\begin{equation}}
\newcommand \ee {\end{equation}}
\newcommand \bea {\begin{eqnarray}}
\newcommand \eea {\end{eqnarray}}
\newcommand \nn {\nonumber}
\newcommand \ve {\varepsilon}

\title{Coarsening of Precipitation Patterns in a Moving Reaction-Diffusion Front}

\author{A. Volford$^{1}$, I. Lagzi$^{2,3}$, F. Moln\'ar Jr.$^{4}$, and
Z. R\'acz$^{4}$}
\affiliation{${}^1$Department of Physics, University of Technology and Economics, 
1521 Budapest, Hungary}
\affiliation{${}^2$Department of Meteorology, E\"otv\"os University, 1117 Budapest, Hungary}
\affiliation{${}^3$Department of Chemical and Biological Engineering, Northwestern University, Evanston, Illinois 60208, USA}
\affiliation{${}^4$Institute for Theoretical Physics--HAS,
E\"otv\"os University, 1117 Budapest, Hungary}

\date{\today}

\begin{abstract}
Precipitation patterns emerging in a 2D moving 
front are investigated on the example of
NaOH diffusing into a gel containing AlCl$_3$. 
The time evolution of the precipitate Al(OH)$_3$ can be observed
since the precipitate redissolves in the excess outer electrolyte NaOH
and thus it exists only in a narrow, optically accessible
region of the reaction front. The patterns display selfsimilar coarsening
with a characteristic length $\xi$ increasing with time as
$\xi(t)\sim \sqrt{t}$. A theory based on
Cahn-Hilliard phase-separation dynamics including dissolution is 
shown to yield agreement with the experiments.
\end{abstract}


\maketitle

Understanding precipitation patterns formed in a moving
reaction front is important both from the basic aspect of
extending our knowledge of phase separation dynamics and from
the point of view of technological applications. Indeed, the
formation of precipitation structures, well localized in space
and time, underlies the notion of the so called "bottom up" designs
\cite{Henisch,lu,grzybowski,tsapatsis}
where one creates structures directly in the bulk.
A natural way of realizing such design
is to have a reaction-diffusion process and control the
dynamics of the reaction zone i.e. control both the position of the
front and the rate of the creation of the reaction product which 
yields the precipitate provided its concentration exceeds a 
threshold value. 

Recently, there have been several attempts at controlling reaction 
zones \cite{grzybowski,giraldo,tsapatsis,Guiding-field} with 
the most promising results emerging from the use of the ionic 
nature of the reagents and realizing control through time-dependent 
electric currents \cite{E-design}. Although these experiments 
demonstrate that one can 
create one dimensional structures reproducibly at the scale 
of $\sim 500\mu$,
the downsizing of the patterns to submicron scales raises several 
problems related to the front. One of them 
is the width of the front which obviously restricts 
the downscaling in the direction of the motion of the front. 
The second one is related to the inhomogeneities within the 
front which may lead to unwanted precipitation structures 
in the plane perpendicular 
to the motion of the front. The scale of these structures clearly limits 
downscaling in the transverse direction. 

The width of reaction fronts has been 
extensively studied theoretically \cite{GR1988,MatPack98} 
as well as experimentally \cite{Kopelman1991,Bazant1999,Tabeling2003} 
and one has some ideas about
the control parameters in this case.
Very little is known, however, about the 
transverse patterns in a moving reaction zone, though
bulk coarsening has been studied in connection with 
the so called gradient-free precipitation experiments
\cite{gradient-free1,gradient-free2}. 

Our aim here is to initiate experimental and 
theoretical studies of the coarsening dynamics of the transverse 
patterns in reaction zones. This task is made feasible experimentally
by overcoming the transparency problem in a way suggested by 
earlier studies of Liesegang type phenomena 
\cite{Zrinyi91,Sultan01,Sultan02,Lagzi07,Volford07,Costello09,Costello09a}.
Namely, the reaction-diffusion process is set in 
a nearly transparent gel and, furthermore, appropriately
chosen electrolytes are used so that the reaction product undergoes 
redissolution in the excess of the outer electrolyte. As a result,
precipitate exists only in a narrow region restricted to 
the reaction zone and
its time evolution can be followed in detail. 

We observed the patterns in the 
moving reaction zone in an experimental setup detailed below.
The visual observations suggested that the system displayed a
selfsimilar coarsening and this was quantified through 
the time dependence of the characteristic lengthscale 
$\xi(t)$ of the pattern, with the main experimental result
being that $\xi (t)\sim \sqrt{t}$.
Theoretically, the effective dimensionality of the coarsening 
system is not entirely obvious, and there are
several candidates for driving the coarsening process.  
We studied this problem by generalizing the Cahn-Hilliard 
theory of precipitation to include sources and sinks 
coming from the emergence of the reaction product in the 
reaction zone, and from the redissolution
of the precipitate in the excess outer electrolyte, respectively. 
The numerical solutions of the equations in three dimension are
in agreement with the experimentally observed $\xi (t)\sim \sqrt{t}$.
This suggests the natural picture that the sources and sinks 
are relevant perturbations on the particle conservation, 
and we observe a curvature driven late-stage coarsening 
in a model with nonconserved order parameter \cite{Binder,Bray}.

In the experiments, a 1 w/w\% agarose (Reanal) gel was prepared with height 
of $0.6-0.7$ cm in a Petri dish. After the gelation process took place ($\approx 2-3$ h), the inner electrolyte (AlCl$_3$) 
was poured on top of the agarose gel to obtain a given concentration in the gel (0.48 mol/L -- 0.56 mol/L). 
After 64 hours the inner electrolyte solution was removed from top of the gel and replaced by the outer electrolyte 
(NaOH of fixed concentration 2.5 mol/L). 

A white precipitation layer formed immediately 
at the gel interface, and this layer started to move into 
the gel. The evolution of precipitation pattern in the moving layer was recorded in reflected light 
using a EOS-20D camera connected to a computer. Typical 
recording period was $t=25$ minutes and 
we analyzed the processes in this time window supposing that the light intensity is proportional to 
the precipitation concentration \cite{Exp-problem}. 

\begin{figure}[htbp]
\includegraphics[width=8.6cm,clip]{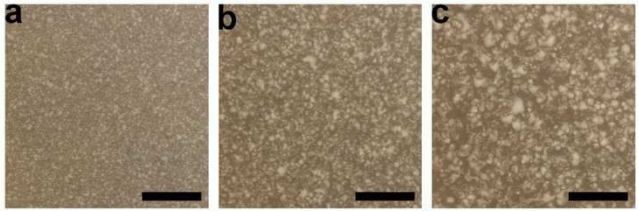}
\caption{Time evolution of the precipitation pattern in the reaction zone
for the samples with [NaOH$]=2.5$M (outer electrolyte) and [AlCl$_3]=0.52$M 
(inner electrolyte). The front moves perpendicularly to the plane of the picture and the
pictures were taken at $t_1=180$s ($a$), $t_2=480$s ($b$), and $t_3=960s$ 
($c$) after the initiation of the reaction. The length of the scalebars 
is 1cm.}
\label{Exp-images_t}
\end{figure}

\begin{figure}[htbp]
\includegraphics[width=8.6cm,clip]{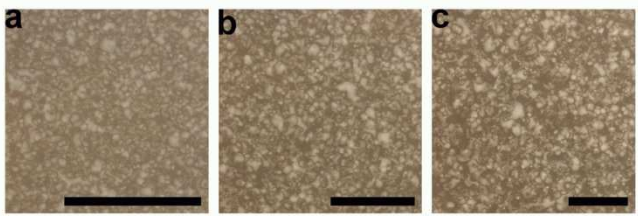}
\caption{Visual demonstration of the scaling in the
evolution seen in Fig.\ref{Exp-images_t}. Parts of 
the panels ($a$) and ($b$) are magnified
by factors $(t_3/t_1)^{1/2}$ and $(t_3/t_2)^{1/2}$, respectively.
The scalebars are the same length of 1cm.}
\label{Exp-images_scale}
\end{figure}

\begin{figure}[htbp]
\includegraphics[width=7.2 cm]{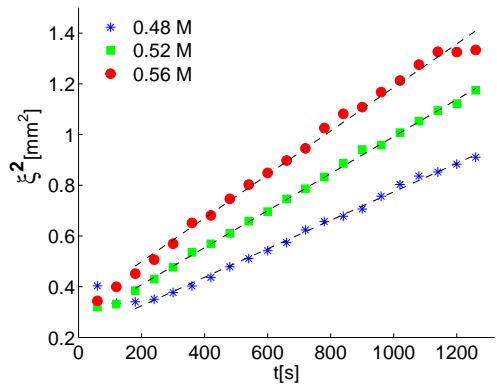}
\caption{Square of the  
characteristic length $\xi^2$ of the precipitation 
patterns plotted as a function of time $t$ for three 
inner electrolyte concentrations and at 
fixed outer electrolyte concentration ($a_0=2.5$M). 
We calculated $\xi^2$ 
from the 2nd moment of the structure factor of the patterns.}
\label{Fig_char_lenght-exp}
\end{figure}

As shown in Fig.\ref{Exp-images_t}, the patterns have a random 
appearance and they coarsen 
with time. 
The coarsening displays selfsimilarity as indicated 
by Fig.\ref{Exp-images_scale} where the pictures have been
magnified by $1/\sqrt{t}$ with $t$ being the elapsed time from the 
initiation of the process. 
The selfsimilarity can be quantified
by measuring the time evolution of the characteristic
length $\xi (t)$ of the precipitation structures. We found $\xi^2$
through calculating the structure factor of the grayscale values
of the pattern and evaluating 
the average of the wavenumber squared 
$\xi^{2}\sim 1/\langle k^2\rangle$. 
The results for three different inner electrolyte concentrations 
are displayed in Fig.\ref{Fig_char_lenght-exp} and we can see 
a well defined diffusive growth $\xi^2\sim 2D_{\perp}t$
regime. 

The transverse diffusion coefficient $D_{\perp}$ can be estimated 
from Fig.\ref{Fig_char_lenght-exp}, and we find 
$D_{\perp}=\xi^2/2t\approx 5\cdot 10^{-10}\quad$ m$^2$s$^{-1}$.
It is remarkable that $D_{\perp}$ is an order of magnitude smaller 
than the diffusion coefficients of the small hydrated ions. 

Our understanding of the observed phenomena is as follows. The
hydroxide ion (outer electrolyte) diffuses into the gel, and the white precipitation layer at the gel interface is produced by the  
reaction with the inner electrolyte (aluminum ions)  
$$
\rm{Al^{3+}(aq) + 3 OH^{-}(aq) \rightarrow Al(OH)_3(s)}\quad .
$$ 
Next, this layer redissolves due to the complex formation 
of aluminum hydroxide in the excess of hydroxide ions producing 
a soluble aluminum complex
$$
\rm{Al(OH)_3(s) + OH^{-}(aq) \rightarrow [Al(OH)_4]^{-}(aq)}\quad .
$$
The combination of the precipitation 
and of the complex-formation processes results in a thin precipitation layer 
moving through the gel. It should be emphasized, however, that the motion 
is solely the transport of chemical species and not of the precipitate.
 
\begin{figure*}[htbp]
\centering
\includegraphics[width=0.325\textwidth,clip]{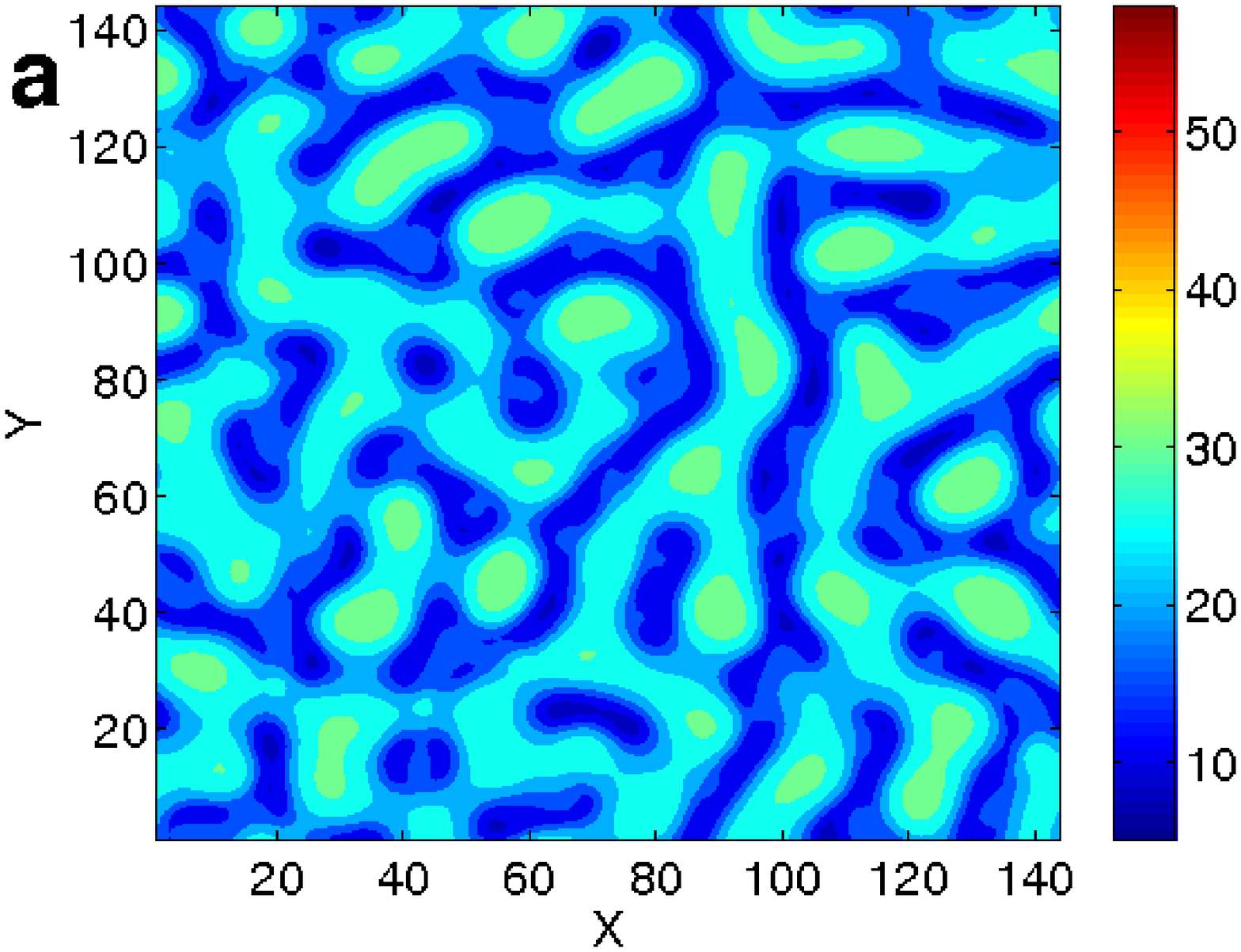}
\includegraphics[width=0.325\textwidth,clip]{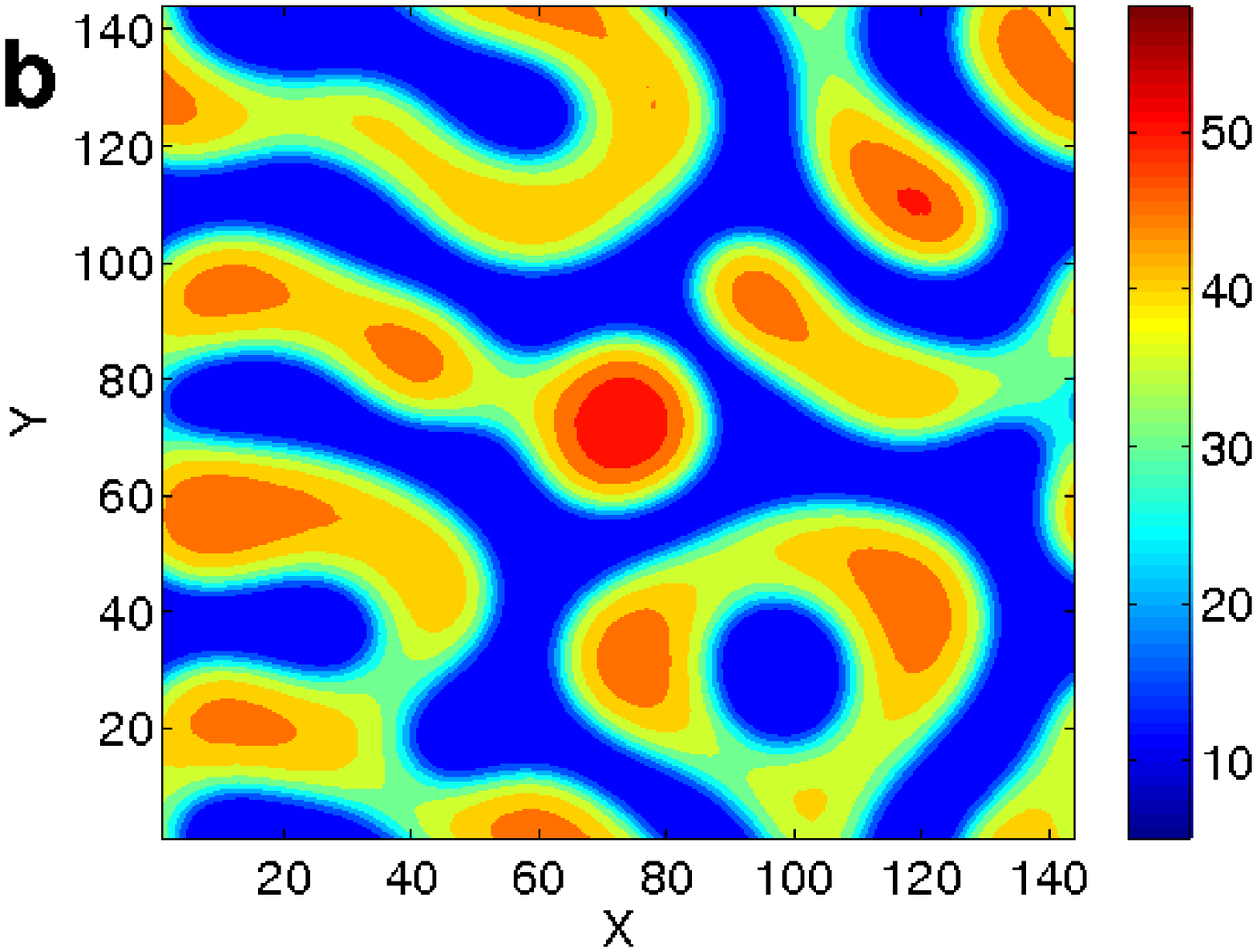}
\includegraphics[width=0.325\textwidth,clip]{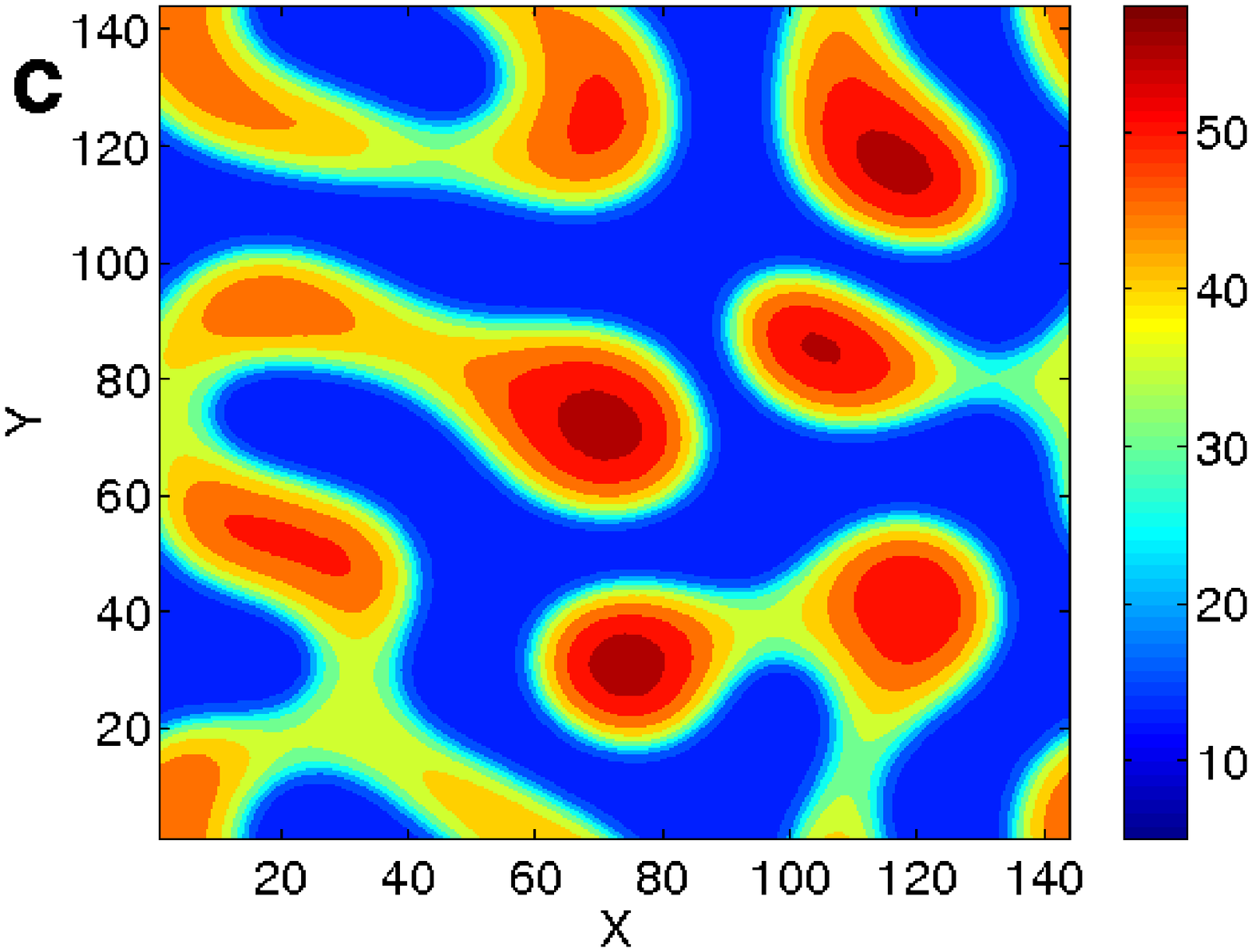}
\caption{Simulation results for the spatial distribution
of C integrated along the $z$ direction. The parameters used were:
$b_0/a_0=0.18$, $c_h/a_0=0.2$, $c_l/a_0=0$, $D=10^{-9}$ m$^{2}$s$^{-1}$, $k_1=8.6$ M$^{-1}$s$^{-1}$, $k_2=0.13$ M$^{-1}$s$^{-1}$, 
$\lambda=0.83\times 10^{-9}$ m$^{2}$s$^{-1}$, $\sigma=10^{-6}$ m$^{2}$, the grid spacing and the time step were 0.01 mm and 0.003 s, respectively. 
The snapshots are taken at $75s$, $600s$ and $1200s$. }
\label{Numsol}
\end{figure*}

Within the above picture, the coarsening in the reaction zone emerges
from a complicated interplay of the dynamics of the reaction zone, 
of the aggregation onto the already present precipitate, and 
of the redissolution process. As a first step in
understanding the interplay of these processes, we approached 
the problem through coarse grained (mean field) equations 
by using a model \cite{ModelB} that has been 
developed in a series of papers during the 
last decade \cite{E-design,Lagzi07}.

The three processes we include are as follows.
First, the reaction of the 
electrolytes $A+B\rightarrow C$ yields the reaction product $C$.
Since the process takes place in a gel, no convection is present 
and it can be modeled as a simple reaction-diffusion process.
This reaction provides the source for the precipitation
which is modeled as a phase separation of $C$s described by
the Cahn-Hilliard equation with a source term.
Finally, the $C$'s redissolution in the excess $A$s 
(complex formation; $A+C\to Complex$) appears as
a sink term both in the Cahn-Hilliard equation and in the 
reaction-diffusion equation for the $A$s. 

The picture can be further simplified if we assume that 
$A+B\to C$ is an irreversible
process for totally dissociated electrolytes 
$A$ and $B$, and that
the case of monovalent ions with equal diffusion coefficients 
are considered. Then the equations take the following form:
\begin{eqnarray}
\partial_t a&=&D\Delta a-k_1ab-k_2ac
\label{e-1}\\
\partial_t b&=&D\Delta b-k_1ab
\label{e-2}\\
\partial_t c&=&-\lambda\Delta(\delta f/\delta c)+k_1ab-k_2ac+
\sqrt c \,\eta
\label{e-3} \, .
\end{eqnarray}
Here $D$ is the diffusion coefficient of the ions, while 
$k_1$ and $k_2$ 
are the rates of reaction and of complex formation, respectively.
We take $k_1$ to be large resulting 
in a reaction zone of negligible width (note that this assumption 
is compatible with the typical reactions used in experimental 
setups that result in Liesegang structures). The thermal fluctuations
in $c$ are described by a noise term $\sqrt{c}\eta$ 
which conserves the total number of $C$ particles \cite{noise}. 
The free energy ($f$) underlying the thermodynamics of 
the phase separation (\ref{e-3}) is assumed to have minima 
at some low ($c_l$) and high ($c_h$) concentrations, and 
it is assumed to be of a Ginzburg-Landau form. Its functional derivative 
most compactly given in terms of a
shifted and rescaled concentration $m=(2c-c_h-c_l)/(c_h-c_l)$
\begin{equation}
\frac{c_h-c_l}{2}\frac{\delta f}{\delta c}= \frac{\delta f}{\delta m} = m - m^3 + \sigma \Delta m\; .
\label{LG}
\end{equation}
Finally, $\sigma$ and $\lambda$ in eq.(\ref{e-3}) 
are setting the spatial- and the 
time scales. They can be chosen to reproduce the correct time and lengthscales in experiments \cite{RZ2000}.

The initial conditions to the above equations are set according 
to the experiments. The outer and inner electrolytes are homogeneously
distributed in the lower ($z<0$) and upper ($z>0$) halfspaces with 
concentrations $a_0$ and $b_0\ll a_0$, while the initial 
concentration of $C$ is $c_0=0$ everywhere. For solutions in finite 
rectangular boxes, we applied no-flux boundary conditions 
at the borders along the $z$ axis while 
periodic boundary conditions were used at the rest of the border planes.

Equations (\ref{e-1}-\ref{e-3}) together with (\ref{LG}) 
were discretized on a uniform 3D grid of size 
144 $\times$ 144 $\times$ 704, and finite-size effects were
checked on grids of size 160$\times$160$\times$568 (no effects were observed) and 96$\times$96$\times$1584 (effects were seen: 
the final characteristic length was increased roughly by factor of  
two).

\begin{figure}[htbp]
\includegraphics[width=7.6 cm]{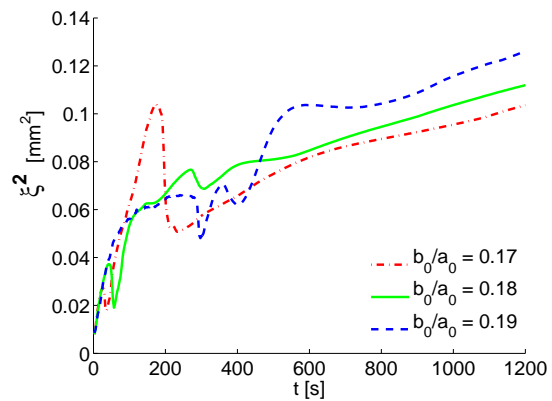}
\caption{Simulation results demonstrating the diffusive growth 
$\xi^2\sim t$ of the characteristic lengthscale $\xi$.}
\label{xi_simulations}
\end{figure}

The resulting ordinary differential equations were integrated 
in time by Euler method (fast simulations were made feasible by 
the parallel programming possibilities of video cards \cite{numerics}).
The results from the simulations are summarized in Figs. \ref{Numsol} 
and \ref{xi_simulations}.

Fig. \ref{Numsol} suggests the existence of two stages in the evolution
of the system. There is an initial period when the unstable, homogeneous
density of $C$s produced by the front begins to evolve towards the 
equilibrium densities (Panel $a$ in Fig. \ref{Numsol}). 
This stage is driven by the initial perturbations (noise) which
determines whether the concentration 
in a given neighborhood grows or diminishes.

The first stage is finished when the high- or low-concentration states 
are reached in a significant fraction of the system and an initial
phase of fast coarsening has also taken already place. 
The emerging high-concentration
regions (Panel $b$ in Fig. \ref{Numsol}) form a structure
which is the base for the next stage, the
self-similar coarsening. It should be noted that the memory of 
this structure can be recognized throughout the
later stages of the coarsening (Panel $c$ in Fig. \ref{Numsol}).

In order to compare with experiments, we calculated 
the structure factor and the characteristic
length $\xi (t)$ by processing pictures
as in case of the experiments.
The results for various inner electrolyte concentrations 
are shown in Fig.\ref{xi_simulations}. 
As we can see, there is a fast initial stage 
governed by noise and short-scale relaxation to the 
equilibrium concentrations, followed by the late-stage 
coarsening. The initial large fluctuations in $\xi$ are 
due to the fact that, at the initial stage, there is a 
significant probability 
for two precipitation bands to coexist. The averaging of 
concentration in the z-direction yields then
an apparent pattern with random correlation length.
Since the Cahn-Hilliard 
equation describes the long-time, long-wavelength properties 
of the coarsening, the short-time fluctuations of $\xi$ may be 
an artificial feature of the results.
Similarly, the explanation of the presence of an 
induction time in the $b_0=0.48$ M experiments may also be outside 
the scope of the CH description. The late stage coarsening,
however, should be correctly given. $\xi(t)$ is indeed 
smooth in this regime and, in agreement with the experiments, 
it shows diffusive behavior ($\xi^2\sim t$).

We should emphasize that the emergence of a 
coarsening state is by no means obvious. The rate of
the arrival of $A$ and $B$ particles to the front decreases 
with time (roughly as $a_0/\sqrt{t}$ and $b_0/\sqrt{t}$)
and the front can advance only in the presence
of surplus $A$s. Thus it is also a conceivable scenario 
that, due to the redissolution process 
($A+C\to Complex$), there is only small concentration of $C$s in the 
front with finite (or perhaps even decreasing) correlation length. 
The selection of a selfsimilar coarsening state is the result of
a delicate interplay between the diffusive advance of the front
and the reaction-diffusion-aggregation processes within the front.

In conclusion, we studied the coarsening of 
precipitation patterns in a thin moving reaction front, and
our experiments suggest that the asymptotic 
dynamics of the system may be interpreted as curvature-driven, 
late-stage coarsening in systems with nonconserved order parameter. 
The theoretical approach, 
based on the Cahn-Hilliard equation coupled to reaction-diffusion 
processes, reproduces all relevant
findings observed in experiments, but the question of
why selfsimilar coarsening is selected by the dynamics is still to
be answered. We believe that, in general, the coupling of 
reaction fronts with phase separation processes opens 
a wide range of possibilities 
for studying new aspects of coarsening and pattern 
formation, and developing this field is of importance for 
submicroscopic technological design.

\acknowledgments
This work has been supported by the Hungarian Academy of Sciences (OTKA No. K 68109, K 68253, and T 72037).


\begin{thebibliography}{99}
\bibitem{Exp-problem}
For $t\gg 25$ minutes, we observed that the coarsening stopped 
and the pattern froze. This 
phenomenon (which is not a finite-size effect since it happens 
irrespectively of the gel thickness and of the concentration 
of the inner electrolyte) appears to be rather complex and 
should be the subject of future studies.
\bibitem{noise} In numerical solutions, the noise
term was realized by exchanging $C$s at neighboring sites proportionally 
to $\sqrt{c}\eta$ where $\eta$ is a random number uniformly distributed 
in the interval $[-0.015,0.015]$.
\bibitem{numerics} The calculations were 
parallelized with CUDA (Compute Unified Device Architecture), 
the new parallel programming architecture 
of NVIDIA video cards. A GeForce 8800GTX video card was used to perform 
computations (2.66 GHz Core2 processor and 4 GB RAM).
\end{thebibliography}
\end{document}